\documentclass[acmsmall,authorversion]{acmart}

\usepackage[hyphens]{url}
\usepackage{hyperref}

\usepackage[english]{babel}
\usepackage{blindtext}
\usepackage{xspace}
\usepackage{subcaption}
\usepackage[absolute,overlay]{textpos}

\newcommand{\TOOL}{{\emph{Priv-Accept}}\xspace}
\newcommand{\BEFORE}{{\emph{Before-Accept}}\xspace}
\newcommand{\AFTER}{{\emph{After-Accept}}\xspace}
\newcommand{\INTERNAL}{{\emph{Additional-Visits}}\xspace}

\setcopyright{none}
\renewcommand\footnotetextcopyrightpermission[1]{} 
\begin{document}
\title[The Internet with Privacy Policies]{The Internet with Privacy Policies: Measuring The Web Upon Consent}

\author{Nikhil Jha}
\affiliation{
\institution{Politecnico di Torino}
\streetaddress{Corso Duca degli Abruzzi, 24}
\city{Torino}
\postcode{10129}
\country{Italy}}
\email{nikhil.jha@polito.it}

\author{Martino Trevisan}
\affiliation{
\institution{Politecnico di Torino}
\streetaddress{Corso Duca degli Abruzzi, 24}
\city{Torino}
\postcode{10129}
\country{Italy}}
\email{martino.trevisan@polito.it}

\author{Luca Vassio}
\affiliation{
\institution{Politecnico di Torino}
\streetaddress{Corso Duca degli Abruzzi, 24}
\city{Torino}
\postcode{10129}
\country{Italy}}
\email{luca.vassio@polito.it}

\author{Marco Mellia}
\affiliation{
\institution{Politecnico di Torino}
\streetaddress{Corso Duca degli Abruzzi, 24}
\city{Torino}
\postcode{10129}
\country{Italy}}
\email{marco.mellia@polito.it}

\begin{abstract}
To protect user privacy, legislators have regulated the use of tracking technologies, mandating the acquisition of users' consent before collecting data. As a result, websites started showing more and more consent management modules -- i.e., Consent Banners -- the visitors have to interact with to access the website content. Since these banners change the content the browser loads, they challenge web measurement collection, primarily to monitor the extent of tracking technologies, but also to measure web performance. If not correctly handled, Consent Banners prevent crawlers from observing the actual content of the websites.

In this paper, we present a comprehensive measurement campaign focusing on popular websites in Europe and the US, visiting both landing and internal pages from different countries around the world. We engineer \TOOL, a Web crawler able to accept the Consent Banners, as most users would do in practice. It lets us compare how webpages change before and after accepting such policies, if present. Our results show that all measurements performed ignoring the Consent Banners offer a biased and partial view of the Web. After accepting the privacy policies, web tracking is far more pervasive, webpages are larger and slower to load.
\end{abstract}





\maketitle
\thispagestyle{empty}

\TPshowboxestrue
\TPMargin{0.3cm}
\begin{textblock*}{14cm}(1.5cm,0.8cm)
\small
\bf
\definecolor{myRed}{rgb}{0.55,0,0}
\color{myRed}
\noindent
Please cite this article as: Nikhil Jha, Martino Trevisan, Luca Vassio, Marco Mellia. The Internet with Privacy Policies: Measuring The Web Upon Consent. ACM Transactions on the Web (2022). DOI: \url{https://doi.org/10.1145/3555352}
\end{textblock*}

\section{Introduction}
\label{sec:intro}

The Web is a complex ecosystem where websites monetize their audience through advertising and data collection. They use Web trackers, i.e., third-party services that collect the visitors browsing history to build per-user profiles and display targeted ads and personalized content~\cite{acar2014web,rizzo2021unveiling,papadogiannakis2021}.
Hundreds of tracking platforms exist, with many of them gathering information from a large base of users and websites~\cite{falahrastegar2014rise,metwalley2015online,pujol2015annoyed,iordanou2018tracing}.

This picture has created tension over online privacy, and regulatory bodies have started governing the scenario. 
Lastly, in May 2018, the EU introduced the General Data Protection Regulation (GDPR)~\cite{gpdr}. It sets strict rules on collecting and storing personal data and mandates firms to ask for informed consent. 
Similarly, the California Consumer Privacy Act of 2018 (CCPA)~\cite{ccpa} gives consumers more control over the personal information that businesses collect. All this has changed the Web too. Nowadays, when users visit a website for the first time, a consent management module -- the commonly called Consent Banner -- prompts, asking the visitors whether they accept the website privacy policy and the use of tracking techniques, and eventually which tracking mechanisms to accept or to block. Upon user's acceptance, the browser activates the accepted tracking techniques and updates the webpage to include all ads and third-party objects.

This challenges the commonly accepted approach to automatically crawl websites to measure the Web ecosystem on privacy~\cite{acar2014web,falahrastegar2014rise,metwalley2015online,englehardt2016online,pujol2015annoyed,iordanou2018tracing,hu2019characterising,rizzo2021unveiling,papadogiannakis2021, vandrevu2019what, pujol2015annoyed, traverso2017benchmark, mazel2019comparison} and performance~\cite{wang2014speedy, de2015http, erman2015towards, bocchi2016measuring, alay2017experience, asrese2019measuring, ruamviboonsuk2017vroom, sivakumar2014parcel, netravali2015mahimahi}. These measurements are typically carried out with headless browsers that access webpages and automatize the collection of metadata and statistics. However, today, these measurements could be biased and unrealistic, with the crawler observing possibly very different content than what a user would get after accepting the privacy policies. In fact, the Consent Banners may hide the actual page content, and the browser may load additional content only after the privacy policy acceptance.
While researchers have shown the importance of carefully choosing which webpages to test~\cite{aqeel2020on}, to the best of our knowledge, we are the first to consider the impact of Consent Banners on automatic measurements.

For this, we engineer \TOOL, a tool to automatically handle the privacy acceptance mechanisms the websites put in place. In a nutshell, \TOOL enables the collection of user-like Web measurements. It overcomes the limitations of traditional crawling approaches, allowing the measurement of the tracking ecosystem to which users are exposed and obtain thus realistic figures on performance. The non-standard way of displaying the Consent Banner, the presence of multiple languages, and the freedom to customize the accept button make automatic detection and acceptance not trivial. We base \TOOL on a keyword list that we thoroughly build to accept the privacy policies automatically. Compared to other solutions~\cite{idontcare,remove,ninja,consentomatic}, \TOOL proves the most robust approach, bypassing the Consent Banner in about $90\%$ of cases when present.

Armed with \TOOL, we run an extensive measurement campaign. We focus primarily on European and US websites that we visit from different countries. We demonstrate how different is the picture we observe before and after accepting the website privacy policies. Interestingly, many websites correctly implement the regulations, activating trackers and personalizing ads only after consent is collected. A researcher collecting statistics by crawling the Web without managing consent could erroneously think that tracking is decreasing with respect to the past~\cite{hu2019characterising}. However, the number of trackers websites embed substantially increases upon acceptance of the privacy policy, in some cases up to 70. As such, popular trackers suddenly become much more pervasive than one can measure using traditional Web crawlers. Similarly, after accepting privacy policies, webpages become more complex and heavier since the browser has to load more objects from more third-party servers. Thus, they are slower to load, so that webpages embedding many trackers and ads double or triple the page load time.

Recently, authors of~\cite{aqeel2020on} showed how important it is to extend the crawling to internal pages. Here, we show that it is also fundamental to correctly handle the Consent Banners when running extensive Web measurements. For this, we offer \TOOL as an open-source tool to incentivize other researchers to contribute. Similarly, we offer all the data we collected for this study and the code to generate the figures to the community in an effort to support reproducibility and encourage other studies.\footnote{\TOOL is available as an open-source GitHub project at: \url{https://github.com/marty90/priv-accept}}

After discussing the scenario and related work in Section~\ref{sec:history}, we present and test \TOOL in Section~\ref{sec:metho}. In Section~\ref{sec:tracking}, we report how different the picture results when checking the Web tracking ecosystem before and after the acceptance of the privacy policies. We then show the implications on performance in Section~\ref{sec:performance}. After discussing Ethics in Section~\ref{sec:ethics} and limitation in Section~\ref{sec:limits}, we summarize our findings in Section~\ref{sec:conclu}.
\section{Background and related work}
\label{sec:history}

Content providers on the Web often monetize the content they offer by using advertisements. To increase their effectiveness, the so-called behavioral advertisement leverages users' interests to provide targeted ads. This is possible thanks to Web trackers, i.e., third-party services embedded in the webpages that gather users' browsing history. Trackers are nowadays largely present on websites and reach the majority of web users~\cite{metwalley2015online,pujol2015annoyed}. Trackers exploit cookies and advanced techniques to enable the collection of personal information~\cite{acar2014web,rizzo2021unveiling,papadogiannakis2021}.


\begin{figure}[t]
    \centering
    \includegraphics[width=0.75\columnwidth]{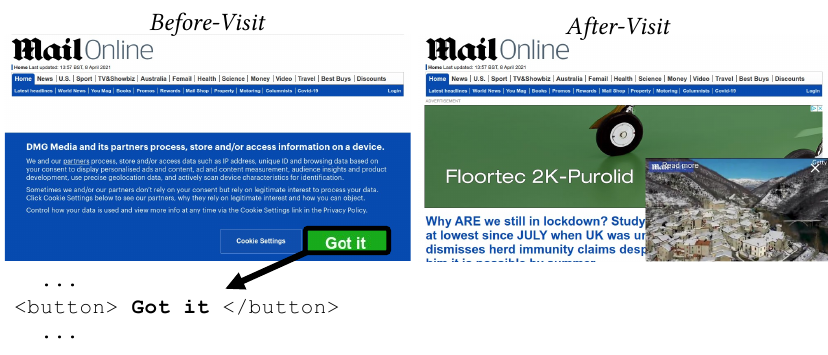}
    \caption{Example of Consent Banner on \texttt{dailymail.co.uk}. Only upon consent, trackers are contacted and ads displayed.}
    \label{fig:cookie_accept_example}
\end{figure}

\subsection{The Role of Legislators}

In this tangled picture, legislators started to regulate the ecosystem to avoid massive indiscriminate tracking that may threaten users' privacy. In 2013, the European Cookie Law~\cite{directive2009} entered into force, which mandates websites to ask for informed consent before using any profiling technology. Later, in May 2018, the General Data Protection Regulation (GDPR)~\cite{gpdr} entered into force in all European member states. It is an extensive regulation on privacy, aiming at protecting users' privacy by imposing strict rules when handling personal information. Unlike previous regulations, it sets severe fines and infringements that could result in a fine of up to €10 million, or 2\% of the firm's worldwide annual revenue, whichever amount is higher. Some websites have already been caught to present legal violations in their Consent Banner implementation~\cite{matte2020cookie} and a large fraction have been shown to use tracking technologies before user consent~\cite{trevisan20194, sanchez2019can}. In the US, the California Consumer Privacy Act (CCPA)~\cite{ccpa} enhances privacy rights and consumer protection for California residents by requiring businesses to give consumers notices about their privacy practices.

As a result, most of the websites now provide explicit Consent Banners~\cite{degeling2018we} and many adopt Consent Management toolsets~\cite{hills2020consent}, making the website content difficult to access until visitors accept the privacy policy. For example, Figure~\ref{fig:cookie_accept_example} shows the same news website homepage before and after accepting the privacy policy. Only upon pressing the ``Got it'' button, the website content is fully loaded and visible.

\begin{figure}[t]
    \centering
    \includegraphics[width=0.5\columnwidth]{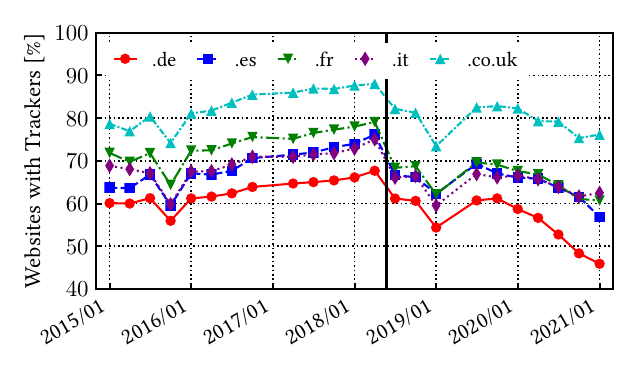}
    \caption{Percentage of websites containing at least one tracker for five European Top-Level domains (from HTTPArchive). The black vertical line indicates the entry into force of the GDPR. Since then, the apparent pervasiveness of tracking decreased.}
    \label{fig:ha_websites_trackers}
\end{figure}

\subsection{The Effect of Consent Banners on Web Measurements}

Despite cases of misuse, the new regulations had a large impact on the web users and complicate the measurement of the tracking ecosystem. A simple Web crawler visiting the websites without accepting the privacy policies would offer a biased picture, with no tracker and no ad being loaded. Hu~\emph{et al.}~\cite{hu2019characterising} already found that the number of third-parties dropped by more than 10\% after GDPR when visiting websites automatically. Conversely, when using a dataset from 15 real users, they measure no significant reduction in long-term numbers of third-party cookies. Dabrowski~\emph{et al.}~\cite{dabrowski2019measuring} draw similar conclusions, finding an apparent decrease in the use of persistent cookies from 2016 to 2018. Sorensen~\emph{et al.}~\cite{sorensen2019before} testify a decreasing trend in the number of third parties during 2018. We quantify this phenomenon in Figure~\ref{fig:ha_websites_trackers}, using the HTTPArchive open dataset~\cite{httparchive}. The curators of this dataset maintain a list of top websites worldwide that they automatically visit using the Google Chrome browser from a US-based server to store a copy of each visited webpage. Using the tracker list detailed in Section~\ref{sec:metho}, we report the percentage of websites embedding one or more trackers for 5 European countries (simply using the Top-Level Domain to identify the country).\footnote{The Top-Level Domain can sometimes be an inaccurate proxy for a website's country. Here, our goal is only to provide a qualitative picture.} We restrict the analysis on those websites that exist for the whole six years-long periods ($9\,196$ website in total).

Figure~\ref{fig:ha_websites_trackers} could suggest that the introduction of the GDPR (the black vertical line in May 2018) results in an abrupt decrease in the number of tracker-embedding websites, a trend that continues up to the moment we write. However, as we will show, these measurements are an artifact due to the GDPR itself. Indeed, the Web crawler used by HTTPArchive can only capture the behavior of the websites as a ``first-time visitor'', before the user accepts any privacy policy. The crawler thus misses third-party trackers and ads.

Research papers that rely on crawling large portions of the Web for different reasons could be affected by the same bias in their measurements. For instance, this would challenge the automatic measurement of the Web ecosystem on privacy~\cite{acar2014web,falahrastegar2014rise,metwalley2015online,pujol2015annoyed,englehardt2016online,iordanou2018tracing,hu2019characterising,rizzo2021unveiling,vandrevu2019what,papadogiannakis2021,aqeel2020on} and counter-measurements~\cite{pujol2015annoyed, traverso2017benchmark, mazel2019comparison}. Moreover, this will also impact those works that rely on crawlers and headless browsers~\cite{avasarala2014selenium} to quantify the impact in the wild of new technologies like SPDY,  HTTP/2~\cite{wang2014speedy,de2015http,bocchi2016measuring,erman2015towards}, 4G/5G~\cite{alay2017experience,asrese2019measuring}, accelerating proxies~\cite{sivakumar2014parcel,wang2016speeding,ruamviboonsuk2017vroom}, or generic benchmark solutions~\cite{netravali2015mahimahi}. At last, even spiders and mirroring tools like Wayback Machine and HTTPArchive may be affected if the website allows the visitor to access its content only after accepting the privacy policy.

\subsection{Related Work and Tools}
\label{sec:related}
Vallina~\emph{et al.}~\cite{vallina2019tales} are the first to consider the impact of the Consent Banner presence. First, they instruct a custom OpenWPM crawler to identify specific Consent Banners, and then they manually verify the results. Unfortunately, they solely focus on the pornographic ecosystem, which they acknowledge to be rather different from the Web at large, and thus their work can hardly generalize.

Recently, authors of~\cite{aqeel2020on} demonstrated that it is fundamental to consider the complexity of the Web ecosystem and include internal pages in every measurement study. They find a number of recent works that neglect internal pages and, as such, might provide biased results. Yet, they ignore the complications due to Consent Banners. Here, we aim at providing an extensive and thorough study of their impact on the Web. Our goal is to enable the automatic study of webpage characteristics as visitors would experience, assuming that most of them accept the default privacy setting as offered by the Consent Banner. Indeed, it has been shown that most users tend to ignore privacy-related notices~\cite{vila2003we, grossklags2007empirical, coventry2016personality}. Considering GDPR Consent Banners, users tend to accept privacy policies when offered a default button via intrusive banners that nudge users ~\cite{CookieBenchmarkStudy,bauer2021you}, which is often the case~\cite{hausner2021dark} with websites presenting large pop-ups or wall-style banners that cover most of the webpage as seen in Figure~\ref{fig:cookie_accept_example}. 

For completeness, notice that cookies are among the simplest tracking mechanisms. Authors of~\cite{papadogiannakis2021} show how practices like cookie synchronization, cookie leaking, and other profiling techniques like canvas fingerprinting are common in today's Web. Similarly, authors of~\cite{jueckstock2021towards} show how the crawling context, in terms of vantage point and browser configuration, has a significant impact on the results. Our work is orthogonal to these to obtain automatic, realistic, reliable and user-centric measurements of the Web.

Focusing on automatic management of Consent Banners, some browser add-ons try to hide them by using a list of CSS selectors of known Consent Banners. The most popular add-ons of this kind are ``I don't care about cookies''~\cite{idontcare} and ``Remove Cookie Banners''~\cite{remove}. Unfortunately, hiding the Consent Banners has an unpredictable behavior, in some cases falling back to privacy policies acceptance, while, in other cases, triggering an opt-out choice. Other proposals, again in the form of browser add-ons, try to explicitly opt-in or opt-out to cookies. For example, ``Ninja Cookie''~\cite{ninja} approves only cookies strictly needed to proceed on the website. Conversely, Autoconsent~\cite{autoconsent} and Consent-O-Matic~\cite{consentomatic} use a set of predefined rules to either opt-in or opt-out to cookies, according to the user configuration. These two are the most similar solutions to \TOOL. However, they are based on a list of actions the browser automatically runs when finding a set of popular Consent Management Platforms (CMPs), limiting their effectiveness. In Section~\ref{sec:ca_vs_com}, we compare \TOOL with Consent-O-Matic -- the most mature tool -- showing that \TOOL offers a much higher coverage. Indeed, the diversity of the Web ecosystem, the presence of multiple languages and the fully customizable choice of Consent Banner buttons make the engineering of \TOOL not trivial.

\section{\TOOL design and testing}
\label{sec:metho}

We explicitly engineer \TOOL to fully automate the visit to websites and collect statistics. The key element of \TOOL is its ability to identify the presence of a Consent Banner and automatically accept privacy policies. We aim at a practical and effective approach to accept privacy policies through the offered button. 

To illustrate \TOOL operation, consider again Figure~\ref{fig:cookie_accept_example}. A large Consent Banner appears on the first visit, and the user shall click on the ``Got it'' button to access the webpage content. \TOOL has to locate this button and click on it automatically. As a result, the website starts loading advertisements and contacting trackers in the background. We refer to these two types of visits as \BEFORE and \AFTER in the remainder of the paper.

We implement \TOOL using the Selenium browser automation tool~\cite{avasarala2014selenium}, the de-facto standard for browser automation, using Google Chrome as browser. Given a target URL, \TOOL carries out the following tasks:
\begin{enumerate}
    \item It navigates to the URL with a fresh browser profile, i.e., with an empty cache and cookie storage. This makes the visit the equivalent of a \BEFORE to the website.
    \item It inspects the Document Object Model (DOM) of the rendered webpage to find a possible \emph{Accept-button} in a Consent Banner. For this, it matches a list of keywords on the text of each node of the DOM. We identify an \emph{Accept-button} if we exactly match any of these keywords. For robustness, we remove leading/trailing/repeated blank characters and the match is performed ignoring the case. We do not use stemming, lemmatization or other techniques for text processing given the specificity of the words to match and the need to support multiple languages.
    \item If \TOOL finds the \emph{Accept-button}, it clicks on the corresponding DOM element (typically a \texttt{<button>}, \texttt{<href>} or \texttt{<span>} element) to accept the privacy policy and logs the success acceptance.
\end{enumerate}

In the beginning, we built \TOOL to look for accept buttons through CSS selectors combined with keywords as done in~\cite{vallina2019tales} and popular add-ons. However, we soon observed that this methodology was too fragile as the use of selectors is strongly CMP-specific and highly customizable by webmasters. The keyword-based approach eases the generalization of the solution. Considering complexity, \TOOL adds marginal overhead to the time required to visit a webpage. Only for very complex webpages, iterating through all DOM elements may require some time, but this is still less than the time needed to load and render the webpage by the browser. 

During each visit, \TOOL stores metadata regarding the whole process in a JSON log file. It includes details on all HTTP transactions and installed cookies. Moreover, it optionally takes screenshots of the webpage during the various phases to allow manual verification.

\TOOL is highly customizable and offers the user various features. It lets the user customize the declared \texttt{User-Agent} and browser language (in the \texttt{Accept-Language} headers). Important to our analysis, it can be configured to run a:
\begin{itemize}
    \item \emph{Warm-up visit}: to populate the browser cache.
    \item \BEFORE: to collect statistics on the webpage before accepting the privacy policy, as a Naive Crawler would do.
    \item \AFTER: to collect statistics on the webpage as it appears after accepting the privacy policy (if an \emph{Accept-button} is found).
    \item \INTERNAL: to a number of webpages of the same website, randomly choosing among the internal links.\footnote{We define internal links as those having the same Fully Qualified Domain Name as the visited website.} We visit internal pages both if \TOOL finds the \emph{Accept-button} and if it does not.
\end{itemize}

For each page visit, \TOOL collect several metadata. Considering QoE metrics, here we focus on the Page Load Time, or \emph{OnLoad} time~\cite{da2018narrowing}. It allows us to compare the webpage rendering performance with and without privacy policy acceptance. It is simpler and faster to compute than the SpeedIndex~\cite{speedindex}, allowing large scale measurements. Notice that we neglect metrics that are not affected by the presence of a Consent Banner, such as the Time-to-first-byte (TTFB). 

Notice that the \AFTER visit can only occur with a warm browser cache in real cases since the browser would have first to complete the \BEFORE visit.
To fairly compare a \BEFORE and \AFTER, in our experiments we run a preliminary \emph{Warm-up visit} before the \BEFORE to fill the browser cache. This lets us appreciate the eventual extra time to load additional components and fairly compare the \emph{OnLoad} on the two visits with the hot cache. Alternatively, \TOOL can erase the HTTP cache and clean the socket pool upon each visit to compare webpage performance with a cold cache. 

\TOOL follows possible redirects during the visits and cases when a script triggers a reload of the webpage. This allows us to manage cases in which the consent banner is hosted on a separate specific landing page than the actual website home page. At last, to limit the impact of random delay due to webpage download and rendering, \TOOL uses quite conservative timeouts before eventually abort the visit. In detail, the DOM inspection starts 5 seconds after the \emph{OnLoad} event. While this clearly slows down the visit of multiple webpages, it maximizes the success rate.

To allow large-scale measurement campaigns, we containerize \TOOL using the Docker container engine~\cite{docker}. In the containerized version, we use Google Chrome version 89 in headless mode and force it to use a standard \texttt{User-Agent} instead of the pre-defined \texttt{ChromeHeadless}.\footnote{The containerized version is available on Docker Hub as \emph{martino90/priv-accept}.}

\subsection{Keyword Selection and Validation}
\label{sec:keywords}

At the core of \TOOL there is the list of keywords to be matched against the webpage content to localize the clickable DOM element for accepting the privacy policy. We thoroughly build this list manually in an iterative way. To handle different languages, we build a list that includes keywords for each country we are interested in. For this work, we focus on 5 European countries, namely France, Germany, Italy, Spain, UK\footnote{In January 2021 UK has enforced the UK GDPR, with practically identical requirements.}, plus the US -- which we use as an example of a large, extra-EU country were privacy laws are in force. For each country, we pick the most popular websites according to the Similarweb lists~\cite{similarweb}, a website-ranking service analogous to Alexa.

\subsubsection{First Round - keyword extraction from top websites}

In the first round, for each of the $5$ countries, we consider the top-200 websites that have a Consent Banner. We randomly choose half of these websites and manually visit them (from Europe) to extract the accept keyword. In total, we visit $500$ websites and identify $186$ unique keywords. We next instruct \TOOL to visit the other half of websites and let it accept the privacy policy, if found. For those where it fails ($233$ cases), we manually visit them to check i) if they have a Consent Banner, and ii) eventually to extract new keywords. With this, we identify $36$ new keywords, $222$ in total. During these steps, we also check that the tool correctly accepts the policy.

\subsubsection{Second Round - testing and keyword increase}

\begin{figure}[t]
    \centering
    \includegraphics[width=0.5\columnwidth]{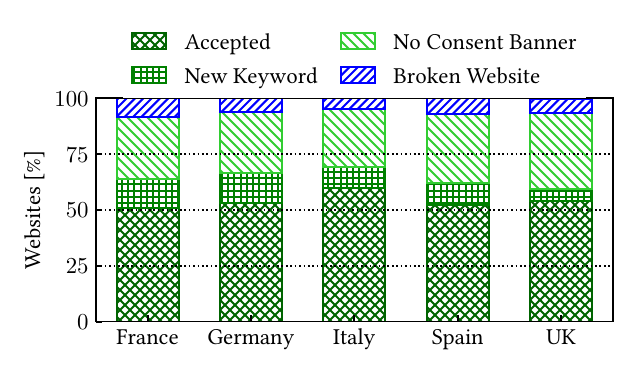}
    \caption{Validation results of \TOOL over 200 randomly picked websites per country. Upon two rounds of keyword selection, \TOOL 92\%-95\% accurate. }
    \label{fig:validation}
\end{figure}

\begin{figure}[t]
    \centering
    \includegraphics[width=0.5\columnwidth]{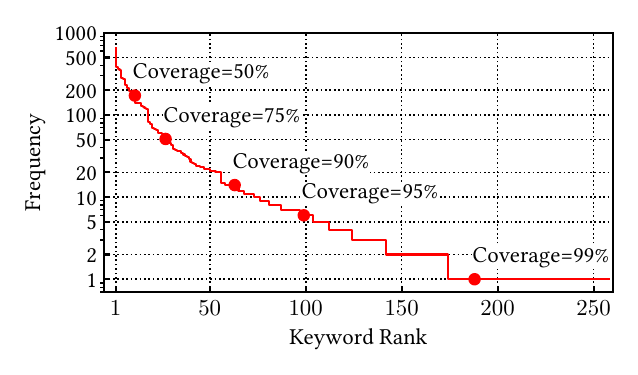}
    \caption{Frequency of the \TOOL keywords, with indication of the coverage at different points. The top-98 keywords already cover 95\% of websites.}
    \label{fig:keywords}
\end{figure}

To evaluate the accuracy of \TOOL in the wild, we next consider $200$ new random websites for each country from the Similarweb lists, $1000$ websites in total. We let \TOOL visit them and manually check the subset of $448$ websites for which \TOOL did not find (and accepted) a privacy policy. We depict the results in Figure~\ref{fig:validation}. \TOOL can accept the privacy policy in more than half of websites, independently from the language. In $6-14\%$ of cases, we find 36 new keywords -- that we promptly add to our list. Interestingly, we find a non-negligible portion of websites ($26-30\%$) that do not present any Consent Banner. At last, \TOOL fails to accept privacy in only $5-8\%$ of cases. Investigating further, this is due to some non-standard behavior of the webpage when accessed in headless mode. For instance, some websites present a CAPTCHA when they detect an automated visit; other websites return a blank webpage. This is a common problem for any crawler-based measurement study~\cite{vastel2020fp}. For completeness, cases of \emph{False Positives} -- i.e., \TOOL clicking on a wrong DOM element -- are possible, although we have not observed any in our manual validation tests. 

At the end of the keyword list building phases, we collect a total of $258 (186+36+36)$ keywords obtained by manually visiting $1181 (500+233+448)$ websites, covering 6 languages.\footnote{In Spain, some websites are in Catalan, rather than in Spanish.} In Figure~\ref{fig:keywords}, we show the distribution of keyword appearance frequency across the entire set of $12\,277$ Similarweb websites (see Section~\ref{sec:dataset} for details on this list). The most common keyword is the string ``Ok''. Red dots indicate the portion of websites covered by the top-$N$ keywords -- i.e., the coverage of the top-$N$ words. The top keywords are very common (note the logarithmic scale on the $y$-axis), with the top-$10$ that cover half of the websites. The top-$98$ keywords cover $95\%$ of the websites, while the remaining appear less than $10$ times each in the whole website set. Clearly, we expect the list of keywords to naturally grow as the tail of the Figure~\ref{fig:keywords} suggests. Notice indeed that more than 80 keywords have been found on a single website. Curiously, we find complex strings like ``I'm fine with this'' or ``Alle auswählen, weiterlesen und unsere arbeit unterstützen''.\footnote{Which translates to ``Select all, keep reading and support our work''.}

\subsection{\TOOL vs. Consent-O-Matic}
\label{sec:ca_vs_com}

\begin{figure}[t]
    \centering
    \includegraphics[width=0.5\columnwidth]{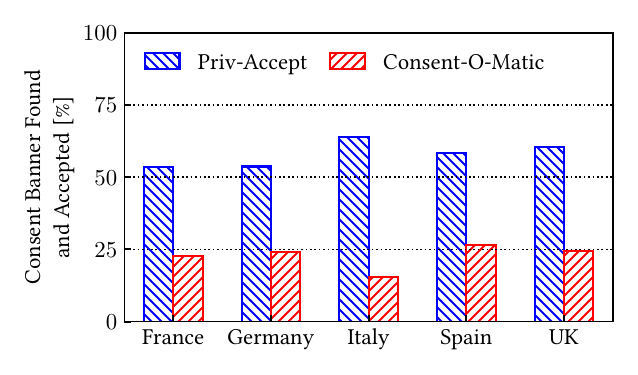}
    \caption{Privacy policy acceptance rate of \TOOL and Consent-O-Matic on 100 websites per country. \TOOL can find and accept Consent Banners on twice as many websites as Consent-O-Matic.}
    \label{fig:ca_vs_com}
\end{figure}

We compare the effectiveness of \TOOL with Consent-O-Matic, the most mature browser plugin designed to offer/deny consent to privacy policies automatically. Unlike our tool, Consent-O-Matic exploits the presence of popular Consent Management Platforms (CMP), services that take care of the management of users' choices on behalf of the website. At the time of writing, Consent-O-Matic allows managing Consent Banners for 35 CMPs. To gauge its performance, we visit the top-100 most popular websites with a Consent Banner for the 5 countries using a Chrome browser with the Consent-O-Matic plugin enabled. Consent-O-Matic accepts the privacy policies in less than 35\% of websites with Consent Banner, and as little as 17\% and 20\% for websites in Italy and UK, respectively. Here \TOOL accepts the privacy policies on all websites by construction.

We then run a second experiment considering another set of 100 websites randomly picked from the Similarweb per country lists. We visit each website with \TOOL and a Consent-O-Matic-enabled browser. Figure~\ref{fig:ca_vs_com} summarizes the comparison. \TOOL accepts the privacy policies in more than 50\% of websites, more than twice the success rate of Consent-O-Matic. These results are in line with those of Figure~\ref{fig:validation}. The remaining websites may not have a Consent Banner, fail to load, or use an unknown keyword. This testifies that the customization of Consent Banners makes it difficult to engineer a generic and simple solution. The keyword-based strategy results more robust than the CMP-based approach (with similar complexity in curating the lists).

\subsection{Dataset and Tracker list}
\label{sec:dataset}

In the following, we use \TOOL to check the impact of using \TOOL when doing large web measurement experiments. We targets a large set of websites popular in France, Germany, Italy, Spain and US, using a test server located in our university campus. For each of the $6$ countries, we use the Similarweb lists to select the top-100 websites from 24 different categories -- see Figure~\ref{fig:ca_category}. These are the top-level unique categories listed in the Similarweb page~\cite{similarwebcategories}. In total, we include $12\,277$ unique websites to visit (as the lists in different countries partially overlap). When visiting websites of a given country, we set the \texttt{Accept-Language} header to indicate the appropriate locale and country language. This behavior can be configured in the \TOOL configuration to allow further experimentation.

We run \TOOL on a single high-end server running 16 parallel instances to speed up the crawl. We instrument it to run a \emph{test sequence}, which consists in a \emph{Warm-up visit}, \BEFORE and \AFTER to the landing page, followed by \INTERNAL to 5 randomly chosen internal pages -- previous studies indeed show that internal and landing pages have different properties~\cite{aqeel2020on}. For each website, we repeat the test sequence $5$ times, randomizing the order of websites to visit in each repetition.  Our main experimental campaign took place for two weeks on April 2021.

We run additional measurement campaigns to investigate specific aspects. To understand whether Consent Banners appear or have a different impact depending on the visitor location, we repeat the above experiments using servers located in the US, Brazil and Japan. We use Amazon Web Services to deploy on-demand servers on the desired availability zone. Here, we aim to check if websites behave differently based on the location of the visitors. Since we are using cloud servers, targeted websites may wrongly recognise the test machines as not regular users and located them in a generic or wrong country. While we cannot check this, we verified that the two most popular commercial IP location databases (IP2Location\footnote{\url{https://www.ip2location.com/}} and MaxMind\footnote{\url{https://www.maxmind.com/}}) map the IP addresses of our crawlers to the correct country. 

To offer a view on a larger number of websites, we visit the top-100\,000 websites according to the Tranco list~\cite{pochat2018tranco}. Unfortunately, the Tranco list does not offer a per-category and per-country rank. We run two separate test sequences: with warm caches, doing (i) \emph{Warm-up visit}, (ii) \BEFORE, and (iii) \AFTER. And with cold caches, (i) \BEFORE, (ii) erase HTTP cache and clean socket pool and (iii) \AFTER. Following this procedure, we ensure a fair comparison between \BEFORE and \AFTER in the two scenarios. Recall that \TOOL allows one to generate any combination of test sequence with warm/cold cache.

To observe how the presence of trackers changes, we rely on publicly-available lists provided by Whotracksme~\cite{whotracksme} (a tracking-related open-data provider), EasyPrivacy~\cite{easyprivacy} (one of the lists at the core of AdBlock tracker-blocking strategy) and AdGuard~\cite{adguard} (a popular ad-blocking tool). For robustness, we merge the three lists and consider as a potential tracker any third-party domain that appear in at least two lists. In total, we obtain $1\,497$ domains that we consider tracking services.\footnote{In the following, we identify them with their \emph{second-level domain name} -- i.e., a hostname truncated after the second label. We handle the case of two-label country code TLDs such as \texttt{co.uk}.} We finally record the presence of a tracker during a visit if the webpage embeds an object from a tracking domain, and the latter installs a cookie with a lifetime longer than one month~\cite{trevisan20194} -- commonly referred to as \textit{profiling cookie}. As such, we divide the HTTP transactions carried out during a visit in: 
\begin{itemize}
    \item First-Party: objects from the same domain of the target webpage.
    \item Third-Party: objects from a different domain than the target webpage.
    \item Trackers: objects from a Third-Party that is a tracking domain and sets a profiling cookie.
\end{itemize}
\section{Impact on Tracking}
\label{sec:tracking}

In this section, we characterize how the Web tracking ecosystem changes if observed with or without accepting the privacy policies. We break down results by Third-Party/Tracker, by country and website category. 


\subsection{Third-Party and Tracker Pervasiveness}

\begin{figure}
    \centering
    \includegraphics[width=0.6\columnwidth]{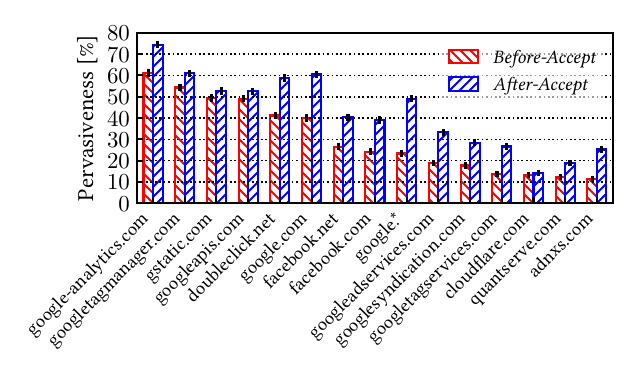}
    \caption{Pervasiveness of the top-15 Third-Parties (percentage of sites they are in) on 10\,542 websites popular in Europe. Most of them are far more pervasive on the \AFTER. 95\% confidence intervals are reported on each bar. }
    \label{fig:ca_prevasiveness_top}
\end{figure}

\begin{figure}
    \centering
    \includegraphics[width=0.5\columnwidth]{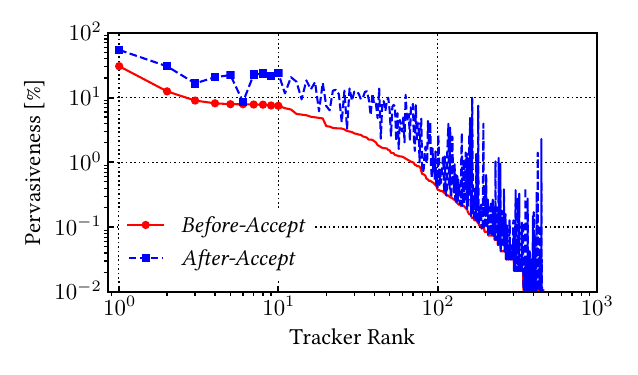}
    \caption{Pervasiveness of the 342 identified Trackers (percentage of sites they are in) in 10\,542 websites popular in Europe. Note that the figure has log-log axes to better show the large variability of Tracker popularity. Also unpopular Trackers result more pervasive on the \AFTER.}
    \label{fig:ca_prevasiveness_all}

\end{figure}

We first study the pervasiveness of Third-Parties and Trackers and check how it varies when we measure it in a \BEFORE or \AFTER. \TOOL found and accepted a Consent Banner on $63.2\%$ of websites. Here, we aim at quantifying the impact of privacy policy acceptance on European websites ($10\,542$ in total) and we exclude those websites exclusively popular in the~US.

We first detail the top-15 most pervasive Third-Parties in Figure~\ref{fig:ca_prevasiveness_top}. The GDPR mandates to obtain informed consent before starting to collect any personal data. As such, Third-Parties may be seen as possibly offending services if activated before accepting the privacy policy.\footnote{Here, we do not enter into the debate of what can be considered a Tracker.} With little surprise, the most pervasive Third-Party is \texttt{google-analytics.com}. It grows from $61\%$ to $74\%$ in popularity on the \AFTER. This value is surprisingly similar to what Metwalley~\emph{et al.}~\cite{metwalley2016using} found in 2016, when they found \texttt{google-analytics.com} appearing in 71\% of websites. The growth is also sizeable for other Google services such as \texttt{googleadservices.com} and \texttt{googlesyndication.com}. Conversely, domains belonging to Content Delivery Networks, such as \texttt{cloudflare.com} and \texttt{cloudflare.net} do not increase their pervasiveness on the \AFTER, likely being not included in the mechanisms of Consent Banners. Interestingly, only 3 out of the top-15 Third-Parties are Trackers -- i.e., present in our tracker list and setting a persistent cookie. \texttt{doubleclick.net} and \texttt{facebook.com} are the most popular ones, with pervasiveness growing from $41\%$ to $58\%$ and from $24\%$ to $39\%$ on the \AFTER, respectively. They are present in more than twice the number of websites than their first competitor (\texttt{quantserve.com}).
In Figure~\ref{fig:ca_prevasiveness_top}, we also report 95\% confidence intervals. It results that the sample proportion (in percentage) of pervasiveness of Third-Parties is an unbiased estimator of the probability $p$ of a Bernoulli random variable. Therefore, by repeating a number of occurrences of a Bernoulli random variable equal to the number of samples, we obtain the number of successes of a binomial random variable. The confidence intervals become the classical binomial proportion confidence intervals. For the sake of completeness, we report error bars also in the following plots. Note, that, given the large number of samples, the confidence intervals are very narrow and not overlapping between \BEFORE and \AFTER, except for the case of \texttt{cloudflare.com}.

Focusing now on Trackers only, we show their pervasiveness in Figure~\ref{fig:ca_prevasiveness_all}. We count $342$ of them. The red curve shows the pervasiveness on the \BEFORE, which is what a naive crawler would report. The blue curve shows how the figure changes on the \AFTER. The Trackers on the $x$-axis are sorted in descending order according to their pervasiveness on the \BEFORE -- hence the \BEFORE curve is monotonically decreasing, while the \AFTER is not. The increase in pervasiveness is general and includes both popular and infrequent Trackers, reaching one order of magnitude in a some cases. On the \AFTER, the number of Trackers that are present on $1\%$ or more of websites grows from $40$ to $90$.
Here, the Spearman's rank correlation is $0.90$, indicating that the Tracker popularity order is approximately the same before and after the privacy policy acceptance. The difference is that their pervasiveness increases.

As it emerges from Figure~\ref{fig:ca_prevasiveness_all}, many Trackers are widespread even on the \BEFORE. This hints at a possibly wrong implementation of the GPDR regulation, which mandates acquiring the visitor's explicit consent before activating any tracking mechanisms. To be precise, the presence of Trackers on the \BEFORE does not necessarily entail a violation of the law. An analysis of the most popular cookies reveals the presence of test cookies during the \BEFORE using a form similar to \texttt{test\_cookie = CheckForPermission}. Google Analytics is a notable example. These cookies are just a check for the possibility of installing profiling cookies upon the user's acceptance. It is thus possible that the \BEFORE pervasiveness of some Trackers includes cases in which only test cookies are actually used (curiously with expiration date longer than a month). Here we limit to observe that often Trackers set some (potentially) profiling cookies even on the \BEFORE.

\textbf{Take away:} 
\textit{Collecting measurements with or without consent to privacy policies leads to a largely different picture. Upon consent, Trackers are far more pervasive than it appears beforehand. \TOOL is instrumental for this goal, thanks to its ability to handle Consent Banners and accept website privacy policies.}

\subsection{Breakdown on Websites}

We now detail the impact of accepting privacy policies on the number of Trackers found in each website, breaking down our results by country and website category.

\subsubsection{Analysis by country}
\label{sec:trackers_country}

\begin{figure*}
    \centering
    \includegraphics[width=0.185\textwidth]{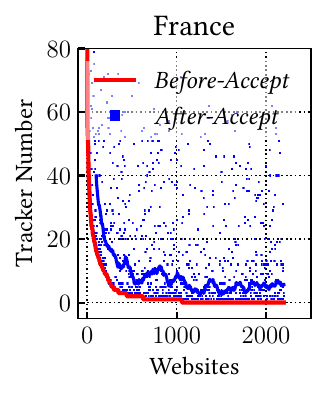}
    \includegraphics[width=0.15\textwidth]{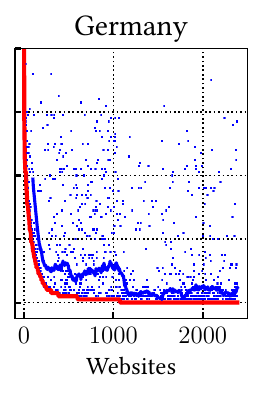}
    \includegraphics[width=0.15\textwidth]{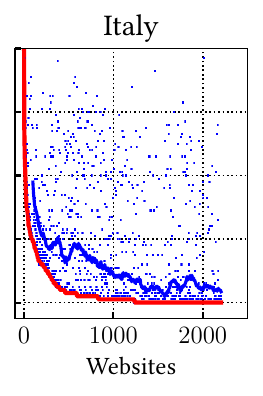}
    \includegraphics[width=0.15\textwidth]{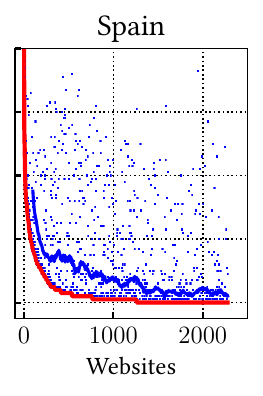}
    \includegraphics[width=0.15\textwidth]{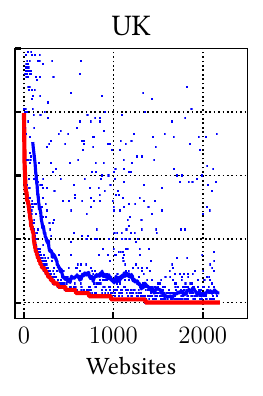}
    \includegraphics[width=0.15\textwidth]{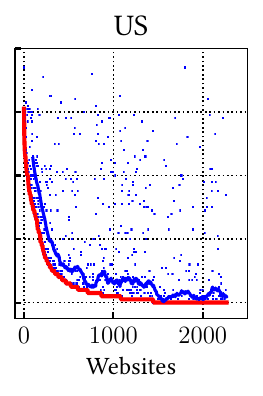}
	\caption{Trackers per website seen on the landing page. Websites (top 2\,500 per country) are sorted by Tracker number on the \BEFORE (red curve). The blue points report the number of Trackers in the \AFTER for the same websites considered in the red curve, while the blue line represent a moving average with a 100-website window.}
	\label{fig:ca_countries}
\end{figure*}

\begin{figure}
    \centering
    \begin{subfigure}[t]{0.495\columnwidth}
        \includegraphics[width=\columnwidth]{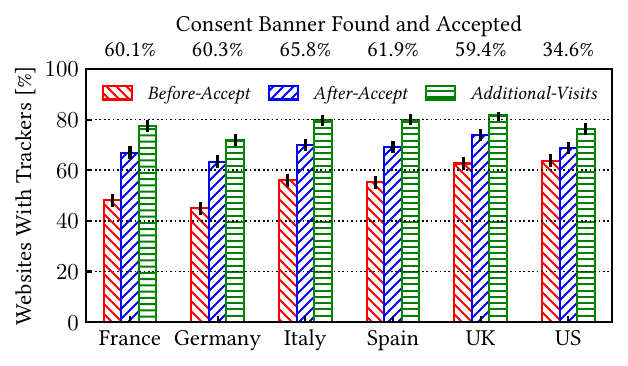}
        \caption{Percentage of websites embedding Trackers. The top $x$-axis details the fraction of websites in such category where \TOOL found and accepted privacy policies.}
        \label{fig:ca_country_one}
    \end{subfigure}
    \begin{subfigure}[t]{0.495\columnwidth}
        \includegraphics[width=\columnwidth]{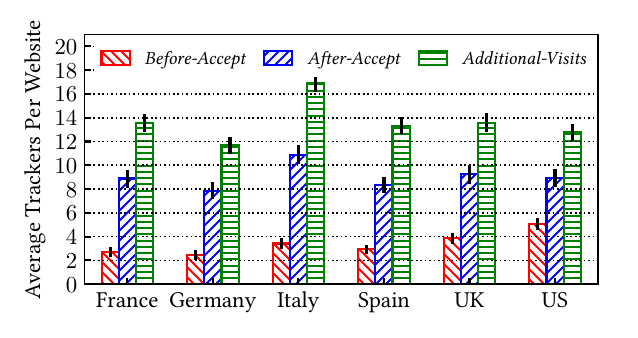}
        \caption{Average number of Trackers per website.}
        \label{fig:ca_country_avg}
    \end{subfigure}
        \begin{subfigure}[t]{0.495\columnwidth}
        \includegraphics[width=\columnwidth]{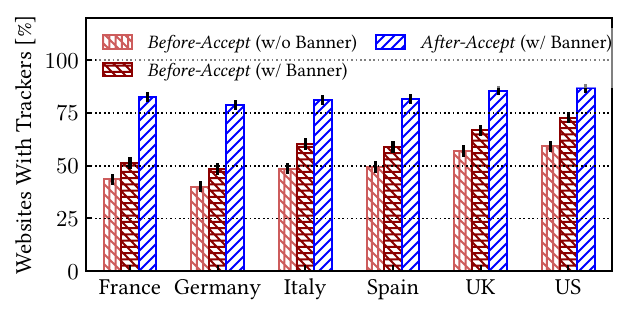}
        \caption{Percentage of websites embedding Trackers, splitting websites with and without a Consent Banner.}
        \label{fig:ca_country_one_sep}
    \end{subfigure}
        \begin{subfigure}[t]{0.495\columnwidth}
        \includegraphics[width=\columnwidth]{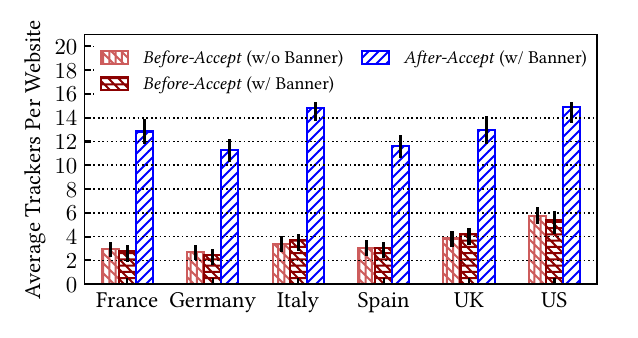}
        \caption{Average number of Trackers per website, splitting websites with and without a Consent Banner.}
        \label{fig:ca_country_avg_sep}
    \end{subfigure}
	\caption{Tracker penetration during different phases of a browsing sessions (top 2\,500 websites per country).  95\% confidence intervals are reported on each bar. On the \AFTER and \INTERNAL, we find many more Trackers.}
	\label{fig:ca_country}
\end{figure}

Figure~\ref{fig:ca_countries} shows websites sorted in descending order by the number of contacted Trackers as measured in the \BEFORE (red curve). This number tends to grow on the \AFTER (blue points), where we observe some websites that present 50-70 more Trackers. To increase readability, in Figure~\ref{fig:ca_countries}, the blue line reports the moving average (with a 100 window) of the number of contacted Trackers on the \AFTER. Curiously, some websites that already include Trackers in the \BEFORE include more Trackers in the \AFTER. This again may hint at a wrong implementation of the Consent Banner, which fails to hinder the presence of offending Trackers. The increase is less remarkable for US-popular websites -- mainly due to the less widespread presence of Consent Banners. 

To better quantify Tracker presence, we show the fraction of websites containing at least one Tracker in Figure~\ref{fig:ca_country_one}. As in Figure \ref{fig:ca_prevasiveness_top}, we report 95\% confidence interval on these sample proportions. About $50\%$ of websites popular in European countries already include at least one Tracker on \BEFORE. This happens more frequently in the UK ($63\%$) and less often in Germany ($44\%$). Again, note that a website embedding a Tracker on the \BEFORE does not necessarily represent a violation of the GDPR, even if this can often be the case~\cite{trevisan20194}. Interestingly, in the US this figure is higher than in European countries. Recalling that the probability of encountering a Consent Banner in the US is lower, this hints at a positive effect of the GDPR on popular European websites. The percentage of websites containing Trackers in the \AFTER grows for all European countries from a $+11\%$ increase in the UK to $+20\%$ for Germany. Confidence intervals never overlap. This increase is moderate ($+5\%$) in the US, given the lower fraction of those websites having a Consent Banner. We complete this analysis by reporting how this fraction increases when performing 5 \INTERNAL as recommended in~\cite{aqeel2020on}. Our results confirm this need, with the chance to observe at least one Tracker that further grows by $5\%$-$10\%$ in \INTERNAL when compared to the \AFTER. Note that, considering each country, none of the confidence intervals overlap between \BEFORE and \AFTER and between \AFTER and \INTERNAL.

We next investigate the quantity of Trackers contacted while visiting websites in Figure~\ref{fig:ca_country_avg}, which shows the average number of Trackers contacted on the websites, separately by country. Also in this case we report 95\% confidence intervals. The sample mean is an unbiased estimator of the true mean, and we can derive confidence intervals through central limit theorem. For all countries, the average amount of Trackers more than doubles on the \AFTER, and performing \INTERNAL further increases this figure (with non-overlapping confidence intervals). In Italy, for instance, this figure grows by a factor of $4$ when comparing \BEFORE and \INTERNAL. As previously noted, the behavior of US-popular websites differs from the European: before acceptance, the number of Trackers is already higher than in popular European websites, while it is comparable after. This hints that popular websites in the United States may be less receptive to GDPR indications. On the opposite side, German-popular websites appear to be the most observant of the regulations, installing Trackers only upon accepting the privacy policies. Afterwards, they reach levels comparable to the other countries. In summary, European websites use the same quantity of Trackers as US ones, although they are often contacted only after accepting the privacy policy. 

To appreciate the variation in the number of Trackers for those websites implementing a Consent Banner, we deepen the analysis by showing separately websites for which \TOOL has found (or not) a Consent Banner. Our goal is to show how Tracker number varies on the \BEFORE and \AFTER for those websites implementing the Consent Banner. Figure~\ref{fig:ca_country_one_sep} shows the percentage of websites with at least one Tracker, and Figure~\ref{fig:ca_country_avg_sep} shows the number of Trackers per website. The dark red bars and blue bars show results on the \BEFORE and \AFTER for those websites where \TOOL \emph{found} a Consent Banner.  As before, the increase of Trackers is sizeable. For completeness, the light red bars report the same measure for those websites where \TOOL \emph{did not} find any Consent Banner.

We finally observe that the probabilistic nature of Web tracking and bidding mechanisms results in a different number of Trackers contacted at each visit. To obtain the most reliable measurements, we test each website $5$ times, each time visiting $5$ internal pages. We note that measuring the fraction of websites containing at least one Tracker (as in Figure~\ref{fig:ca_country_one}) is moderately impacted by the number of tests. Indeed, when considering a single \AFTER per website, overall, we find $69.1\%$ of them containing one (or more) Trackers. Repeating $5$ times the test and considering whether we find at least one Tracker among all visits, this percentage increases only to $70.0\%$. Similarly, the average number of Trackers (as in Figure~\ref{fig:ca_country_avg}), increases from $6.5$ to $7.8$. We report additional details on this in the Appendix and in Figure~\ref{fig:visit_nb}.

\subsubsection{Analysis by category}

\begin{figure*}
    \centering
    \begin{subfigure}[t]{\textwidth}
        \includegraphics[width=\textwidth]{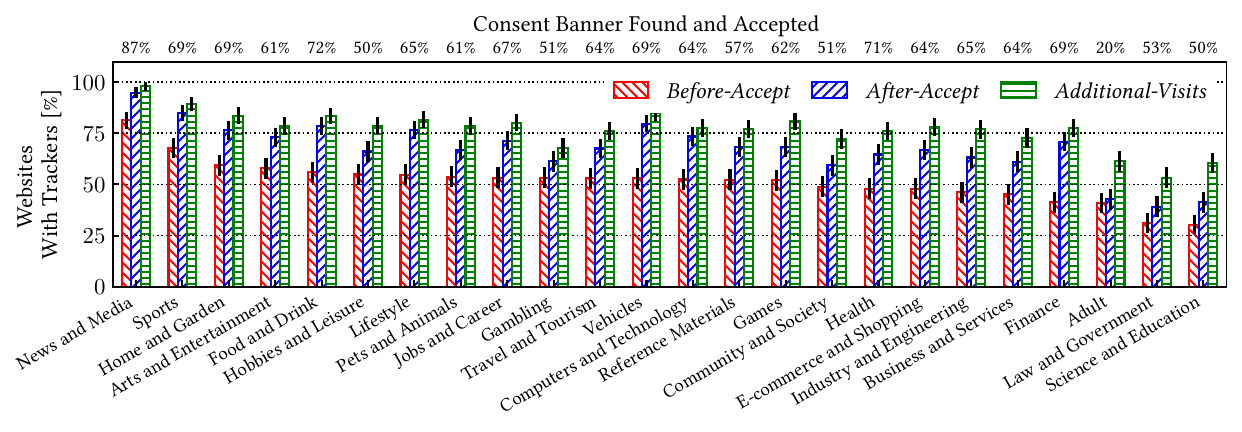}
        \caption{Percentage of websites embedding Trackers. The top $x$-axis details the fraction of websites in such category where \TOOL found and accepted privacy policies.}
        \label{fig:ca_category_one}
    \end{subfigure}
    \begin{subfigure}[t]{\textwidth}
        \includegraphics[width=\textwidth]{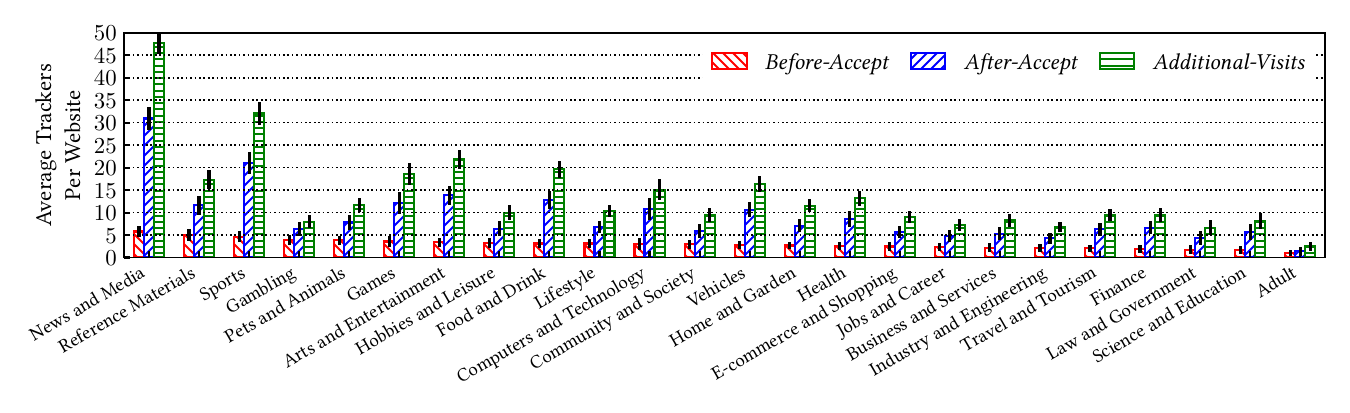}
        \caption{Average number of Trackers per website.}
        \label{fig:ca_category_avg}
    \end{subfigure}
	\caption{Trackers penetration and number on websites (top 2\,500 per country) during different phases of a browsing session, separately by category. We sort categories from the highest to the lowest percentage of websites with Trackers in \BEFORE. 95\% confidence intervals are reported on each bar. In some cases (e.g., News and Media), on the \AFTER and \INTERNAL the increase is very pronounced.}
	\label{fig:ca_category}
\end{figure*}

We now break down the picture by category, showing the results in Figure~\ref{fig:ca_category}. We explicitly target websites of $24$ categories, each containing the top-$100$ websites for the considered countries.

Starting from Figure~\ref{fig:ca_category_one}, we report the percentage of websites of a given category that contain at least one Tracker. As before, there is a large increase from \BEFORE to \AFTER.  Exceptions are the \textit{Adult}, \textit{Law and Government} and \textit{Gambling} categories, where the confidence intervals overlap. For \textit{Adult} this is likely due to the low number of websites with Consent Banners ($20\%$) and confirms the peculiarity of the tracking ecosystem on Adult websites~\cite{vallina2019tales}. As previously observed in Figure~\ref{fig:ca_country_one}, performing \INTERNAL further increases the chance of encountering at least one Tracker, even though in this case the increase is limited and we observe some overlaps between \AFTER and \INTERNAL confidence intervals.

Moving to the number of trackers per website shown in Figure~\ref{fig:ca_category_avg}, we observe large increase in the \AFTER case, confirming that most Trackers appear only after the user accepts the privacy policies and when visiting internal pages. Here, differences across categories are all pronounced, with those categories that heavily depend on advertisements (\textit{News and Media}, \textit{Sports}, \textit{Games}, \textit{Arts and Entertainment}) that have to rely on a large number of Trackers to support behavioral advertisements. This is noticeable already on the \BEFORE. For example, access to a \textit{News} website leads to contact $5.7$ Trackers on average in \BEFORE. Here, \TOOL successfully accepts the privacy policies in $87\%$ of cases. Indeed, being \textit{News} websites very popular, they tend to correctly implement the privacy regulations and to show a well-configured Consent Banner. Upon acceptance, suddenly, the number of Trackers becomes almost 6 times higher ($30.9$ for \textit{News}) and 9 times higher when doing \INTERNAL ($47.7$ trackers on average). For \emph{Sport}, \emph{Food and Drink} and \emph{Arts and Entertainment} the average number of Trackers more than triples in \AFTER. Only for the \textit{Adult} category confidence intervals overlap.

These numbers are particularly interesting if read in the perspective of recent works. Englehardt~\emph{et al.}~\cite{englehardt2016online}, in 2016, measured an average of 35 Trackers per website on News websites. In 2021, we find similar numbers ($30.9$) on the \AFTER, while, due to the spread of Consent Banners, on the \BEFORE we would only find $5.7$, on average. On Sport category, Englehardt~\emph{et al.}~\cite{englehardt2016online} measured $27$ Trackers per website. In 2021, we find $21.0$ on the \AFTER, while only $4.6$ on the \BEFORE. These results well highlight the need for correctly handling the Consent Banners to observe the extensiveness of web tracking. 
In a nutshell, thanks to \TOOL, we obtain the fundamentally different figure in the \AFTER and \INTERNAL. 

The case of \textit{Adult} websites is worth a specific comment. \TOOL finds the Consent Banner on only $20\%$ of them, and a manual check on $50$ of them confirms that the large majority of them do not offer any Consent Banner. Tracking is also limited upon acceptance, and the confidence intervals between \BEFORE and \AFTER even overlap. Similar results were previously found by Vallina~\emph{et al.}~\cite{vallina2019tales}, where the authors suggest that the specialized pornographic advertisement ecosystem may cause this behavior: usually, trackers and advertisers related to pornographic websites do not operate outside of them -- often evading popular tracker lists.

\textbf{Take away:} 
\textit{Upon consent, the number of Trackers embedded in websites increases by a factor of up to 4 times. European and US websites end up with a similar number of Trackers. The increase is particularly pronounced for certain website categories -- for example, News and media or Sport websites -- that rely on ads as revenue stream.}

\subsection{Visits from Outside Europe}

\begin{figure}[t]
    \centering
    \includegraphics[width=0.5\columnwidth]{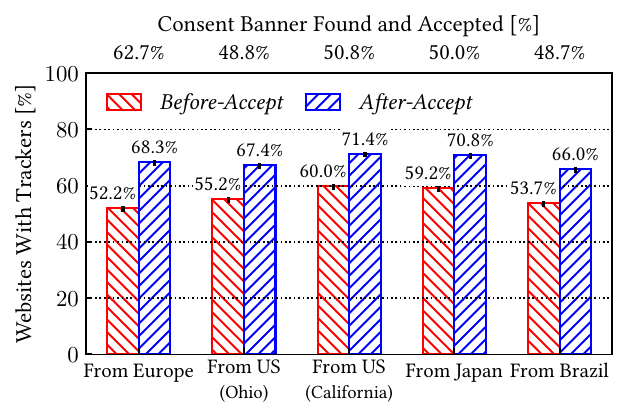}
    \caption{Websites with Trackers (12\,277 from the Similarweb lists) when crawling from different countries. 95\% confidence intervals are reported on each bar. From non-European countries, \TOOL found fewer Consent Banners, but the amount of Trackers on the \AFTER is similar. Outside Europe, top-ranked websites tend to include more Trackers.}
    \label{fig:ca_us}
\end{figure}

We now consider additional measurement campaigns using crawling servers in the Amazon AWS data centers located in the US (Ohio and California), Japan and Brazil. Figure~\ref{fig:ca_us} summarizes our findings. First, notice how \TOOL accepted privacy policies on around $10\%$ fewer websites (about $1\,150-1\,200$) when run from outside Europe, as reported on top $x$-labels. 
Checking the screenshot taken by \TOOL during the visit on a random subset of these websites, we confirm that no Consent Banner is displayed. We can conclude that some websites customize the Consent Banners based on visitors' properties, such as their location. If the visit comes from not EU country, no Consent Banner is shown.

This different behaviour of websites affects also the statistics of the fraction of websites that embed trackers in the \BEFORE and \AFTER visits. Visiting from outside Europe leads to an increase of Tracking on the \BEFORE in all cases, while, on the \AFTER, changes are limited.

\textbf{Take away:} 
\textit{The crawling location location has some impact on the results. This is mostly due to websites that show or not show the Consent Banner based on the user's location, thus not enabling or enabling tracking on the \BEFORE.}

\begin{figure}[!t]
    \centering
    \begin{subfigure}[t]{0.495\columnwidth}
        \includegraphics[width=1.0\columnwidth]{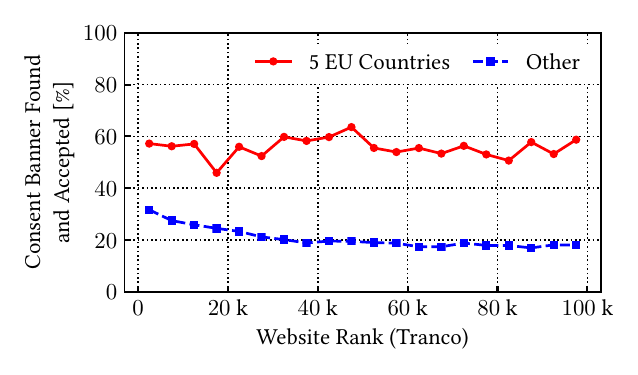}
        \caption{Percentage of websites with a Consent Banner.}
        \label{fig:tranco_rank}
    \end{subfigure}
    \begin{subfigure}[t]{0.495\columnwidth}
        \includegraphics[width=1.0\columnwidth]{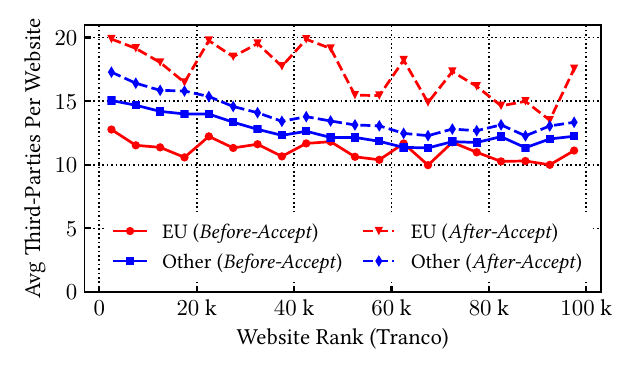}
        \caption{Average number of Third-Parties per website.}
        \label{fig:tranco_tp}
    \end{subfigure}    
     \caption{Percentage of websites with a Consent Banner and average Third-Parties per website over the top-100 k websites in Tranco list, computed every $5\,000$ websites in the rank. Top websites are more likely to implement a Consent Banner in a \TOOL supported language.}
    \label{fig:tranco}   
\end{figure}

\section{Impact on Complexity and Performance on Top-100k Websites}
\label{sec:performance}

In this section, we measure the impact of accepting privacy policies on the webpage characteristics and loading performance. Trackers and Third-Party objects that the browser has to load and display upon consent may impact the amount of data to download and the rendering performance. Here, we do not restrict on a per-country or per-category analysis and use the crawl on the top-$100\,000$ websites according to the Tranco global list. 

For each website, we visit only the landing page, doing a \emph{Warm-up} visit to fill the browser cache, followed by a \BEFORE and \AFTER. We compare results on the latter two visits, considering only those websites for which \TOOL successfully accepted the privacy policy, which happens on $23$\% of websites. This is in line with the previous findings, as the Tranco list is a worldwide rank and includes (i) European websites in a language different from those for which we built the keyword list and (ii) websites based in non-European countries for which regulations do not apply. To give more insights, we detail the percentage of websites with a Consent Banner on the Tranco list in Figure~\ref{fig:tranco_rank}, computed every $5\,000$ websites in the rank. The solid red line reports the percentage for websites popular in the 5 European countries we target. Websites belong to this set if (i) they appear in the Similarweb ranks for the 5 countries or (ii) the Top-Level Domain belongs to the 5 countries.\footnote{The Tranco list does not provide a per-country rank.} Out of these $6\,917$ websites, \TOOL accepts the privacy policy on $3\,861$ ($55.8$\%), which is close to what we have obtained with the Similarweb ranks ($54.7\%$). This percentage does not change with website popularity. Conversely, for the remaining websites (blue dashed line), the share of websites where \TOOL found a Consent Banner is $32$\% for the top-ranked and then it settles around $20$\%, hinting that some globally popular websites tend to implement a Consent Banner even if they are based outside Europe, using a language supported by \TOOL (likely English). In 2020, Hills~\emph{et al.}~\cite{hills2020consent} found that popular CMPs are present on almost $10$\% of websites in the top-10 k Tranco list. Here, with \TOOL, we can affirm that Consent Banners (regardless the employed CMP) appear in more than $30$\% for the same set of websites.

The high number of Consent Banners found for the 5 European countries reflects in a large increase of the number of Third-Parties from the \BEFORE to the \AFTER, as shown in Figure~\ref{fig:tranco_tp}. The solid red line highlights that these websites already include, on average, $11.1$ Third-Parties in the \BEFORE. In the \AFTER, the average grows to $17.3$. Differently, the increase for the non-EU websites is smaller -- see the area between the blue solid and dashed lines. In the \BEFORE, Third-Parties are larger than for the 5 European countries if we compare the solid blue and red lines. This is due to the larger presence of non-EU websites, which do not have to implement a Consent Banner. In the \AFTER (dashed blue line), the increase is moderate, not reaching the values of the 5 European countries (red dashed line), potentially because \TOOL misses many \emph{Accept-button} in non-supported languages and of possible custom tracking domains not present in our lists. For the sake of completeness, in the Appendix, we report the same picture as in Figure~\ref{fig:tranco_tp} showing the number of Trackers instead of Third-Parties, providing similar insights.

\textbf{Take away:} 
\textit{For the five European countries considered, the percentage of websites with a Consent Banner (and the number of third parties) is approximately flat with respect to website rank. For the websites of the remaining countries, \TOOL may miss some \emph{Accept-button} due to the usage of local languages.}

\subsection{Impact on Page Objects and Size}

\begin{figure}[!t]
    \centering
    \begin{subfigure}[t]{0.495\columnwidth}
        \includegraphics[width=\textwidth]{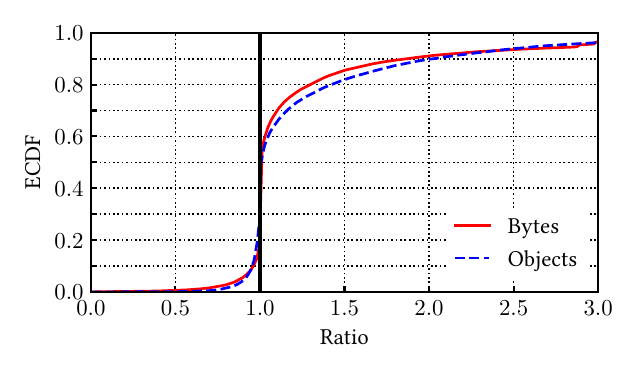}
        \caption{Distribution of the page size (in bytes and objects) ratio over all websites.}
        \label{fig:ca_perf_size}
    \end{subfigure}
    \begin{subfigure}[t]{0.495\columnwidth}
        \includegraphics[width=\textwidth]{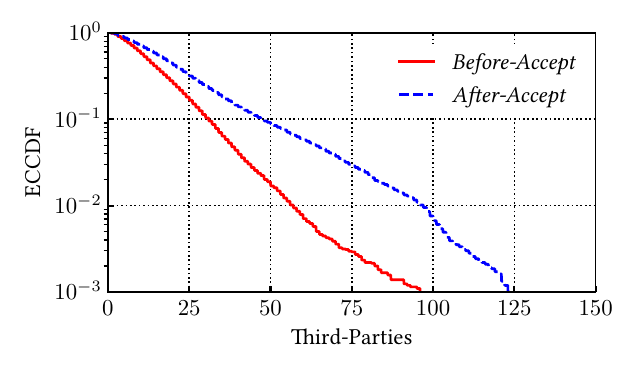}
        \caption{Distribution of the number of Third Parties. Notice the log scales.}
        \label{fig:ca_perf_tp}
    \end{subfigure}
	\caption{Webpage characteristic before and upon consent to privacy policies (Tranco list). On the \AFTER webpages are larger and include more Third Parties.}
	\label{fig:ca_perf}
\end{figure}

We focus on the webpage complexity in terms of  bytes and objects to download. We compute the ratio $R$ between the measurement on the \BEFORE and \AFTER, i.e., $R = x_{\textit{After}}/x_{\textit{Before}}$, where $x$ is the metric of interest. We show the results in Figure~\ref{fig:ca_perf_size}, separately for total downloaded bytes and objects. As expected, accepting the privacy policy increases the webpage size ($R>1$) by a sizeable factor. For instance, about $9$\% of websites download more than twice the objects, and about $5$\% of websites sees an increase of 3 times or more.

Interestingly, we also observe some websites that are lighter in the \AFTER than in the \BEFORE. Investigating further, these cases are mostly due to the lack of additional content upon acceptance coupled with the saving of not loading the CMP objects on the \AFTER. This happens commonly on those websites that either add a Consent Banner despite not using tracking mechanisms, or that contact Trackers and Third-Parties even before the user has accepted the privacy policies. While the former might be seen as an excess of caution, the latter cases are likely violating the privacy regulations.

To better characterize the differences, we quantify the number of Third-Parties seen in the \BEFORE and \AFTER. We show the Empirical Complementary Cumulative Distribution Function (ECCDF) in Figure~\ref{fig:ca_perf_tp}. On median, websites rely on $12$ Third-Parties on the \BEFORE. This figure grows to $17$ on the \AFTER. The ECCDF highlights the tail of the distribution where we observe those websites that rely on a very large number of Third-Parties: the percentage of websites with more than $50$ grows from $1.8\%$ to $9.2\%$, with $3.0\%$ including more than 75 Third-Parties upon acceptance. This growth in the number of Third-Parties is mostly due to an increase of Trackers and objects related to advertisements that gets loaded after accepting the privacy policy. 
We also perform statistical tests to compare whether the mean and median of the two sample distributions are statistically different at level 0.05 (t-Test for the mean and Mood’s test for the median). Both result statistically significant in \AFTER. 
In Appendix, we include the picture as above, plotting the number of Trackers instead of Third-Parties, leading to similar conclusions.

\textbf{Take away:} 
\textit{When the Consent Banner is accepted, websites are larger, with $9$\% of them containing more than twice as many objects. Websites including more than $50$ Third-Parties increase from $1.8\%$ to $9.2\%$.}


\begin{figure}[!t]
    \centering
    \begin{subfigure}[t]{0.495\columnwidth}
        \includegraphics[width=\textwidth]{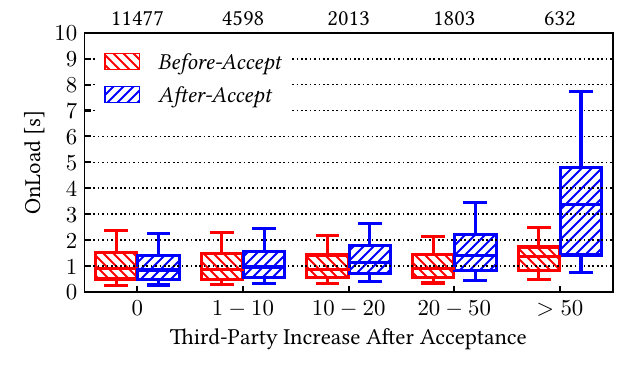}
        \caption{Warm Browser Cache.}
        \label{fig:ca_onload_warm}
    \end{subfigure}
    \begin{subfigure}[t]{0.495\columnwidth}
        \includegraphics[width=\textwidth]{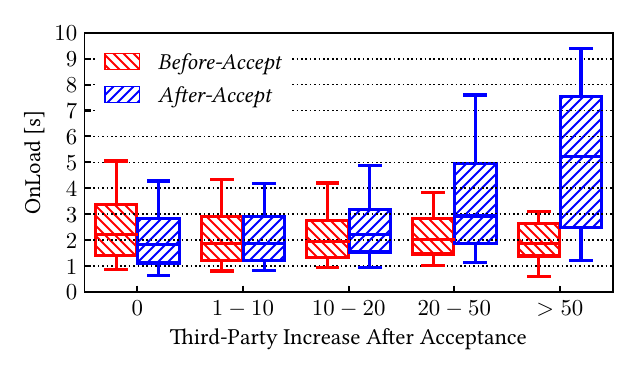}
        \caption{Cold Browser Cache.}
        \label{fig:ca_onload_cold}
    \end{subfigure}
    \caption{OnLoad time of websites versus the increase of Third-Party number upon acceptance (Tranco list). The cardinality of each category is reported on the top axis of the left-most figure. Website adding many Third Parties on the \AFTER are also slower to load. }
     \label{fig:ca_onload}
\end{figure}

\subsection{Impact on Page Load Time}

The Third-Party domains appearing after acceptance are generally devoted to advertisements, analytics and Web tracking. 
Contacting them has direct implications on the page load time and, indirectly, on the users' QoE~\cite{da2018narrowing}. We thus expect this to cause an increase on the page load time because the browser has to resolve the server name via DNS and contact more servers. For instance, this ultimately limits the advantages offered by new protocols like the stream multiplexing and the header compression offered by HTTP/2 and HTTP/3.

To gauge this, we dissect the webpage load time in Figure~\ref{fig:ca_onload}, comparing separately visits with a warm cache (Figure~\ref{fig:ca_onload_warm}) and with a cold cache (Figure~\ref{fig:ca_onload_cold}). In case of warm cache, we run a \emph{Warm-up visit}, then the \BEFORE and \AFTER. In case of of cold cache, we run the \BEFORE without a \emph{Warm-up visit}. Then we erase the HTTP cache and socket pool, then we run the \AFTER.

We report the distributions of the \textit{onLoad} time for websites with similar number of additional Third-Parties that are loaded in the \AFTER. We use boxplots, where the boxes span from the first to the third quartile and whiskers from the $10^{th}$ to $90^{th}$ percentile. The central stroke represents the median. The number of websites in each set is detailed on the top the respective boxplot. As expected, the more Third-Parties are loaded upon acceptance, the larger the time needed to load the webpage and the larger its variability. Especially for the websites that add more than 10 Third-Parties, the distributions are remarkably different on the \BEFORE and \AFTER. Considering visits with cold browser cache (Figure~\ref{fig:ca_onload_warm}), those website with $20-50$ additional Third-Parties, the median \textit{onLoad} time passes from $0.91$ to $1.41$ seconds. The difference increases for the $632$ websites adding more than $50$ Third-Parties upon acceptance. Here, the median \textit{onLoad} time increases from $1.35$ to $3.38$ seconds, more than doubling. Notice also the tail of $25\%$ of websites loading in more than $4.8$\,seconds, which happens in less than $2\%$ of cases during the \BEFORE. We already observed such an increase in our previous study~\cite{traverso2017benchmark}, where we measured that median \textit{onLoad} time increases by $1.3s$ when cookies policies are accepted. 
We statistically compare all these couples of sample distributions between  \BEFORE and \AFTER, testing differences in the median at a significance level 0.05 (Mood's test). The test is passed in all cases, showing statistically significant differences.

Similar considerations hold for visits with a cold browser cache (Figure~\ref{fig:ca_onload_cold}). As expected, with the clean cache, websites load time increases -- compare values in Figures~\ref{fig:ca_onload_warm} and~\ref{fig:ca_onload_cold}. Those that do not add new Third-Parties tend to load slightly faster on the \AFTER, potentially due to the absence of the Consent Banner. In fact, differences are statistically significant in the median of the distributions between \BEFORE and \AFTER,  except for the group $1-10$ additional Third-Parties.
Again, we observe that those adding several Third-Parties after acceptance have much higher \textit{onLoad} time on the \AFTER than on the \BEFORE:  The median \textit{onLoad} time increases from $1.8$ to $5.2$ seconds. Finally, we observe that the \textit{onLoad} time values tend to be lower than what measured in older works, potentially because of the advances of content delivery network and increased hardware and software performance. Bocchi~\emph{et al.}~\cite{bocchi2016measuring} measured a median \textit{onLoad} time of 3s in 2016 on a similar albeit smaller set of websites.

\textbf{Take away:} 
\textit{Measuring the webpage load time of websites without considering the implications of accepting the Consent Banners would result in a very biased measurement. Websites that include many more Third-Parties upon acceptance are significantly slower to load.}
\section{Ethical considerations}
\label{sec:ethics}

During our measurements, we took care to avoid harming the crawled webpages. We contacted each website $5$ times in a span of two weeks and accessed a limited number of internal webpages each time. Considering that the target of our analysis were some of the most popular websites in Western countries, our belief is not to have caused an overload on the servers or any undesirable side effect. Moreover, since we did not interact with Third-Parties after accepting the privacy policies -- including displayed ads -- we consider not to have significantly altered the economic ecosystem of the crawled websites. We only used the standard HTTP and HTTPS ports for our measurements, carefully avoiding any type of port scanning procedures, and we used large timers to avoid creating any kind of congestion.

\section{Limitations}
\label{sec:limits}

Our work presents a few limitations, some of which could be addressed in future work.

First, \TOOL is designed to accept the privacy policy in the Consent Banner. It could be interesting to extend \TOOL to consider different keywords to choose the different options (e.g., to Opt-Out) on the Consent Banner and verify if websites correctly implement the end-user choice.

Second, the keyword list is manually compiled and static. We leave for future work the design of an automatic mechanism to enlarge and maintain the list. For instance, one can envision a community effort to enrich the list. It would also be interesting to consider some Natural Language Processing-based approaches to compile the keyword list automatically.

Third, currently, \TOOL uses a global list of keywords, regardless of the website's language. Although unlikely, a keyword may have a different meaning in another language, leading to false positives. A simple solution would be to add support for country- and language-specific lists of keywords.

Considering the results on the web tracking pervasiveness, we here focused on those based on tracking cookies, and we ignore advanced techniques for web tracking such as CNAME cloaking~\cite{dao2021}, a technique to embed Trackers as first-party domains, or device fingerprinting~\cite{rizzo2021unveiling}. Our results are thus an underestimation of the extensiveness. This problem is general and not specific for \TOOL.

Moreover, in our experiments, we set the browser language according to the country of each visited website. However, websites may customize their behaviour depending on the users' language, as some are already doing based on the user's location. \TOOL already allows configuring the content of \texttt{Accept-Language} header, making it possible to study this aspect in detail.

\section{Conclusions}
\label{sec:conclu}

In this paper, we demonstrated how the recent regulations had changed the Web landscape, challenging its automatic measurements through traditional Web crawlers. Websites now massively deploy Consent Banners to obtain visitors' consent for using tracking technologies and collecting personal data. As a result, webpages appear very different once users provide their consent. This has vast implications when measuring Web tracking, webpage characteristics, website performance, and any measurement based on Web crawling.

In this paper, we engineered \TOOL, a tool that automatically crawls websites accepting the privacy policy when a Consent Banner is found. We run it on many websites popular in Europe and worldwide. Our results highlighted how the observed picture of the Web varies when measured upon accepting privacy policies: Web Trackers and Third-Parties suddenly become more pervasive, websites more complex, and slower to load.

We release \TOOL as an open-source project, along with the dataset used throughout the paper. We based it on a set of keywords and, thus, has margins for improvement. We foster its use by the research community to contribute to it and extend our results. We also hope \TOOL will be included as part of the public projects that provide periodic Web measurements. Our goal is to keep developing \TOOL to enrich the keyword list, implement additional functionalities, adding the possibility to deny the privacy policies, a much more complex task. For this, we envision the design of more sophisticated approaches to manage Consent Banners, likely based on recent advances in Natural Language Processing and Machine Learning.

\section*{Acknowledgments}
The research leading to these results has been funded by the European Union's Horizon 2020 research and innovation program under grant agreement No. 871370 (PIMCity project) and the SmartData@PoliTO center for Data Science technologies.

\bibliographystyle{ACM-Reference-Format}
\bibliography{reference}

\newpage

\section*{Appendix}
\label{sec:appendix}

\subsection*{Impact of Repeated Visits on Tracking Measurements}

\begin{figure}[!h]
    \centering
    \includegraphics[width=0.6\columnwidth]{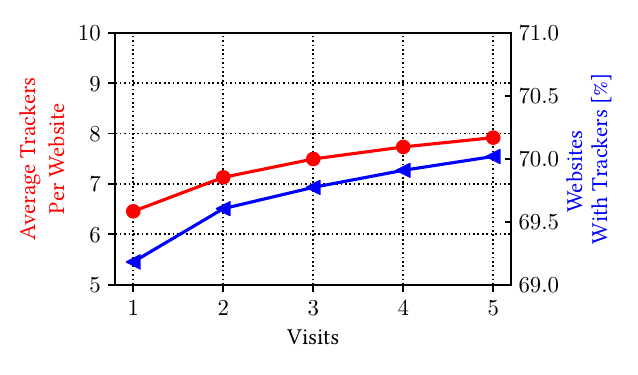}
    \caption{Variation of tracker number with different numbers of repeated visits. Measurements have sizeable despite moderate variation when repeated.}
	\label{fig:visit_nb}
\end{figure}

We here complement the analysis we carried out on the last paragraph of Section~\ref{sec:trackers_country}. Web tracking involves a number of mechanisms ( real-time bidding among all ) that result in the same page containing different Trackers on multiple visits. To obtain a reliable picture, we repeat each test 5 times. In Figure~\ref{fig:visit_nb}, we show how two macroscopic tracking measurements vary with different number of repetited visits for each website. The blue line in the figure shows the fraction of websites that contain at least one Tracker when measured with an increasing number of test repetitions. It is moderately affected by the number of tests, increasing from 69.1\% with a single repetition to 70.0\% with 5 repetitions. Similarly, the average number of Trackers, increases from $6.5$ to $7.8$.

\newpage

\subsection*{Trackers per Website (Tranco List)}

We here report the same analyses depicted in Figure~\ref{fig:tranco_tp} and Figure~\ref{fig:ca_perf_tp} showing the number of Trackers instead of the number of Third-Parties. The two pictures lead to similar conclusions.

\begin{figure}[!h]
    \centering
    \includegraphics[width=0.5\columnwidth]{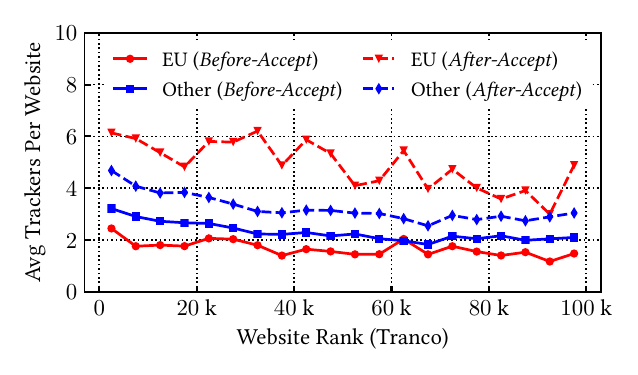}
    \caption{Average number of Trackers per website (Tranco list).}
    \label{fig:tranco_trackers}
\end{figure}

\begin{figure}[!h]
    \centering
    \includegraphics[width=0.5\textwidth]{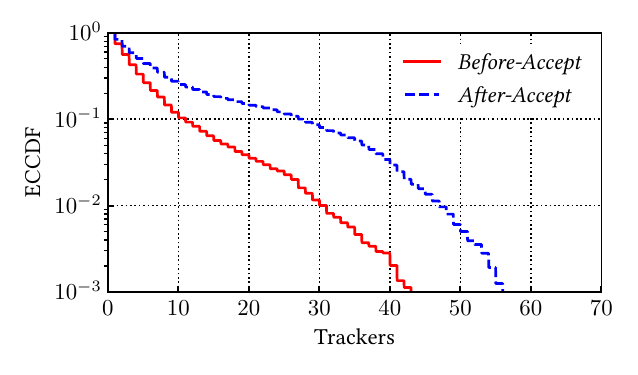}
    \caption{Distribution of the number of Trackers (Tranco list). Notice the log scales.}
    \label{fig:ca_perf_tracker}
\end{figure}

\end{document}